\documentclass[aps,pr,twocolumn,
superscriptaddress,groupedaddress,nofootinbib,floatfix]{revtex4-1}  

\usepackage[colorlinks,citecolor=blue,urlcolor=blue,linkcolor=blue]{hyperref}
\usepackage{soul}
\usepackage{amsfonts}
\usepackage{amssymb}
\usepackage{amsmath}
\usepackage{amsthm}
\usepackage{graphicx}
\usepackage{appendix}
\usepackage{bbold}
\usepackage{dsfont}
\usepackage{enumitem}
\usepackage{xcolor}
\usepackage{yfonts}
\usepackage{MnSymbol}

\usepackage[normalem]{ulem}
\newcommand\redsout{\bgroup\markoverwith{\textcolor{red}{\rule[0.4ex]{3pt}{0.7pt}}}\ULon}

\providecommand{\U}[1]{\protect\rule{.1in}{.1in}}

\usepackage{color}
\usepackage{listings}
\definecolor{darkgreen}{rgb}{0,0.35,0}
\definecolor{Rood}{rgb}{1, 0, 0}

\begin{document}

\title{One-loop Schwinger effect in the presence of Lorentz-violating background fields}

\author{Rafael L.~J.~Costa}
\email{rafaelljc@id.uff.br}
\affiliation{Instituto de F\'isica, Universidade Federal Fluminense, Campus da Praia Vermelha, Av. Litor\^anea s/n, 24210-346, Niter\'oi, RJ, Brazil}
\author{Rodrigo~F.~Sobreiro}
\email{rodrigo\_sobreiro@id.uff.br}
\affiliation{Instituto de F\'isica, Universidade Federal Fluminense, Campus da Praia Vermelha, Av. Litor\^anea s/n, 24210-346, Niter\'oi, RJ, Brazil}

\begin{abstract}
In this work we make use of proper-time method to evaluate Schwinger effect in the presence of Lorentz-violating background fields. Specifically, we evaluate the one-loop effective Lagrangian in the presence of Lorentz-violating time-like vector background $b_\mu$ and pseudo-scalar $m_5$ for the pure electric case. The imaginary part of the effective Lagrangian is computed in order to calculate the Schwinger effect. Comments on phenomenology are also performed.
\end{abstract}

\maketitle

\section{Introduction}

Lorentz symmetry plays an important role in quantum field theory and particle Physics \cite{Bargmann:1946me,Bargmann:1948ck}. However, in the last decades, several works with proposals presenting violation of Lorentz symmetry have received lots of attention, see for instance 
\cite{Kostelecky:1988zi,Carroll:1989vb,Kostelecky:1989jw,Colladay:1996iz,Colladay:1998fq,Jackiw:1999yp,Kostelecky:2003fs,Kostelecky:2005ic,Belich:2004ng,Mariz:2005jh,Gomes:2009ch,Diaz:2011ia,SSantos:2015mzs,Santos:2016bqc,Gomes:2019uay,Ferreira:2020wde,Filho:2021rin,Mariz:2021cik}. Since the effects of such theories are expected to appear only at the Planck scale, some cumulative effects have been studied in order to obtain evidence of Lorentz symmetry breaking \cite{Bertolami:1996cq,Bluhm:1997ci,Bailey:2006fd,Kostelecky:2013rv}. We can also mention a few works trying to find possible corrections that Lorentz symmetry breaking produces at classical level. For example, in classical electrodynamics we refer to \cite{Bailey:2004na,Wu:2006pd,Belich:2003xa,Lehnert:2004be,Altschul:2006zz,Borges:2014ita,Araujo:2020jpx,Silva:2020dli,Silva:2021fzh} and references therein.

Another path in the study of theories with Lorentz violation concern their radiative corrections. In the late 1990s, Roman Jackiw and Alan Kostelecky \cite{Jackiw:1999yp} introduced the idea that it is possible to generate terms with Lorentz symmetry breaking at quantum level. Basically, starting from a Quantum Electrodynamics (QED) Lagrangian with violation of Lorentz and CPT symmetries in the fermionic sector (by radiative corrections and integrating out the fermion fields), new violation terms are generated in the electromagnetic sector. Many other works about such ideas can also be mentioned, \cite{Chen:1999ws,Chung:1999pt,Chung:2001mb,Chaichian:2000eh,Anacleto:2014aha,Carvalho:2018vtr,Ferrari:2019dvu,Ferrari:2021eam,SSantos:2015mzs,Santos:2016bqc}. See also \cite{Santos:2014lfa,Santos:2016dcw} for the non-Abelian case.

Within Lorentz invariant QED, radiative correction is used, for example, to deduce the Euler-Heisenberg (EH) effective Lagrangian, first found by Werner Heisenberg in \cite{Heisenberg:1936nmg}. Some years later, more precisely in 1951, Julian Schwinger achieved the same result by using the proper-time method \cite{Schwinger:1951nm}. The proper-time method was first introduced by Vladimir Fock in \cite{Fock:1937dy} when working with the Dirac equation (and also in Yoichiro Nambu's work \cite{Nambu:1950rs}). Finally, the method was employed by Schwinger for the study of gauge invariance \cite{Schwinger:1951nm}. The proper-time method has advantages because it preserves some formal symmetries during calculations. For example, it has the advantage of being gauge and Lorentz invariant simultaneously.

In addition to Schwinger's analysis of gauge invariance in \cite{Schwinger:1951nm}, he showed a very interesting and non-perturbative effect of QED, the creation of electron-positron (E-P) pairs in vacuum due to a static external electric field. Known as the Schwinger effect, this is an extremely small effect for weak fields, it is expected that it will become measurable for fields on the order of $\mathcal{E}=10^{14}V/cm$ \cite{Valluri:1982ip,dittrich1985effective2}.
In the next years a facility in Shanghai, China, named Station of Extreme Light (SEL), are aimed to produce a laser with $100PW$ of power, willing to be powerful enough to reach the Schwinger limit. Another experiment that aimed to probe the non-perturbative regime of pair creation is the Laser Und European X-ray Free Electron (XFEL) Experiment (LUXE), a new experiment in planning at DESY Hamburg using the electron beam of the XFEL \cite{Jacobs:2021fbg}.

Given that, we may reach the Schwinger limit in the coming years, our main goal in this work is to calculate the Schwinger effect in the presence of Lorentz-violating background fields in the fermionic sector. Our proposal is that if we are able to obtain results for the Schwinger effect with better precision, we may use this as a way to probe the Lorentz symmetry breaking. We find for the Schwinger effect in the presence of Lorentz-violating background fields, in despite of providing extremely small contributions, an increase (or decrease - depending on the magnitude of the background fields) in the pair creation probability. It was also possible to find an estimate of the electric field strength needed to have an effect of Lorentz-violating fields on the probability of E-P pair creation that could be measured in the LUXE experiment, but the intensity is extremely high.

Our work is organized as follows. In the section \ref{GT} we will present the model that will be used and its respective simplifications and approximations. The $1$-loop effective action for pure electric field is also calculated and we show that our result reproduces the generation of the CPT-even term for a time-like background pseudo-vector field and a pseudo-scalar background field. In section \ref{PC}, we use the result obtained in the previous section to calculate the imaginary part of the effective Lagrangian and thus calculate the production rate of E-P pairs. In the section \ref{PHE}, we make some comments regarding phenomenology and present some limits for the Lorentz violating fields. Finally, in section \ref{FINAL} we present the conclusions.

\section{One-loop effective Lagrangian}\label{GT}

Here we will work with a QED extension which involves only Lorentz-violating terms in the fermionic sector. The general Lagrangian of this model is given by\footnote{Natural units, $\hbar=c=1$, are employed in the present work.} \cite{Colladay:1998fq,Kostelecky:2003fs}
\begin{align}
\label{L0}
    \mathcal{L}_{QED_{ex}}=-\frac{1}{4}\mathcal{F}_{\mu\nu}\mathcal{F}^{\mu\nu}+\bar{\psi}\left(\Gamma_{\mu}\Pi^{\mu}-M\right)\psi\, ,
\end{align}
where
\begin{eqnarray}
\label{LVfields}
    \Gamma_{\mu}&\equiv&\gamma_\mu+ c_{\nu\mu}\gamma^{\nu}+d_{\nu\mu}\gamma^{5}\gamma^{\nu}+e^{\mu}+if_{\mu}\gamma^{5}+\frac{1}{2}g_{\alpha\beta\mu}\sigma^{\alpha\beta}\;,\nonumber\\
    M&\equiv&m+ im_{5}\gamma_{5}+a^{\mu}\gamma_{\mu}+b^{\mu}\gamma_{5}\gamma_{\mu}+\frac{1}{2}h^{\mu\nu}\sigma_{\mu\nu}\;.
\end{eqnarray}
The covariant derivative is defined as $\Pi^{\mu}=i\partial^{\mu}-eA^{\mu}$, the field strength is written as $\mathcal{F}_{\mu\nu}=\partial_{\mu}A_{\nu}-\partial_{\nu}A_{\mu}$, $A_\mu$ is the gauge potential, $\psi$ is the fermionic field. Greek indexes range from $0$ to $3$. The parameters $m$ and $e$ stands for electron mass and charge, respectively. All tensor fields in \eqref{LVfields} are Lorentz-violating background fields. They feature spacetime with anisotropy by selecting privileged directions. The tensor fields in the first line of \eqref{LVfields} has no mass dimension while in the second line they have mass dimension $1$. Furthermore, tensor fields with an odd numbers of indexes do not preserve CPT while an even number of indexes preserve CPT\footnote{For the explicit CPT features of the fields see, for instance, \cite{SSantos:2015mzs}.}. For the $\gamma$ matrices, we define them from the Clifford algebra $\left\{ \gamma^{\mu},\gamma^{\nu}\right\}=-2\eta^{\mu\nu}$ with the Minkowski metric given by $\eta_{\mu\nu}=\mathrm{diag}\left( -1,1,1,1\right)$. Moreover,
\begin{eqnarray}
    \sigma^{\mu\nu}&=&\frac{i}{2}\left[\gamma^{\mu},\gamma^{\nu}\right]\,,\nonumber\\
    \gamma_{5}&=&-\frac{i}{4!}\epsilon^{\mu\nu\sigma\beta}\gamma_{\mu}\gamma_{\nu}\gamma_{\sigma}\gamma_{\beta}\,.
\end{eqnarray}

It is important to notice that all Lorentz-violating fields and also the gauge field $A_\mu$ are considered here as background fields. Therefore, the fermionic filed is the only quantum field in \eqref{L0}. In fact, for simplicity purposes, we consider a particular case of \eqref{LVfields}, namely $\Gamma_\mu=\gamma_\mu$ and $M=m+im_5\gamma_5+b^\mu\gamma_5\gamma_\mu$. Moreover, $b^\mu$ is chosen as a timelike vector, $b_{\mu}=\left(b,0,0,0\right)$; and the field $m_5$ is included in order to probe if there are contributions to the Schwinger effect arising from it, even though it can be removed by a chiral transformation \cite{Kostelecky:2013rta}.

Since the main goal of this paper is to obtain the E-P pair creation rate, we need to evaluate the effective action. This task can be done by integrating out the fermion field. We can start from
\begin{align}
\label{GF2}
    Z\left[A,b,m_{5}\right]&=e^{iW^{\left(0\right)}\left[A\right]}\frac{\int\mathcal{D}\bar{\psi}\mathcal{D}\psi e^{i\int d^{4}x\left\{ \bar{\psi}\left(\gamma^{\mu}\Pi_{\mu}-W\right)\psi\right\} }}{\int\mathcal{D}\bar{\psi}\mathcal{D}\psi e^{i\int d^{4}x\left[\bar{\psi}\left(i\gamma^{\mu}\partial_{\mu}-m\right)\psi\right]}}\;.
\end{align}
where $W=\gamma_{5}\left(im_{5}+b\gamma_{0}\right)-m$ and $W^{\left(0\right)}\left[A\right]=-\frac{1}{4}\int d^{4}x\left(\mathcal{F}_{\mu\nu}\mathcal{F}^{\mu\nu}\right)$. Employing the usual Gaussian integral formula for the fermion field, we get
\begin{align}
    \label{GF3}
    Z\left[A,b,m_{5}\right]&=e^{iW^{\left(0\right)}\left[A\right]}\frac{\det\left(\gamma^{\mu}\Pi_{\mu}-W\right)}{\det\left(i\gamma^{\mu}\partial_{\mu}-m\right)}\,,
\end{align}
from where we derive the one-loop effective Lagrangian,
\begin{align}
    \label{L1}
    \mathcal{L}^{\left(1\right)}&=-i\ln \det\left(\gamma^{\mu}\Pi_{\mu}-W\right)+i\ln \det\left(i\gamma^{\mu}\partial_{\mu}-m\right)\,.
\end{align}
Taking advantage from the fact that the determinant of an operator $\Xi$ is invariant under a similarity transformation $S$, as follows, $\text{Det}\left[\Xi\right]=\text{Det}\left[S\Xi S^{-1}\right]=\text{Det}\left[\Xi S\Xi S^{-1}\right]^{\frac{1}{2}}$,
we can choose $S$ to be, for instance, the charge conjugation. Hence,
\begin{align}
    \label{L2}
    \mathcal{L}^{\left(1\right)}&=-\frac{i}{2}\ln\det\mathcal{H}+\frac{i}{2}\ln\det\mathcal{H}_{0}\,,
\end{align}
with
\begin{eqnarray}
    \label{H}
    \mathcal{H}&=&\left[\gamma^{\mu}\Pi_{\mu}-\gamma_{5}\left(im_{5}+b\gamma_{0}\right)+m\right]\left[-\gamma^{\nu}\Pi_{\nu}+\right.\nonumber\\
    &-&\left.\gamma_{5}\left(im_{5}-b\gamma_{0}\right)+m\right]\;,\nonumber\\
    \mathcal{H}_{0}&=&\left(i\gamma^{\mu}\partial_{\mu}-m\right)\left(-i\gamma^{\mu}\partial_{\mu}-m\right)\;,
\end{eqnarray}
and we used the fact that $C\gamma^{\mu}C^{-1}=-\gamma^{\mu}$ and $Cb^{\mu}C^{-1}=b^{\mu}$ and $Cm_{5}C^{-1}=m_{5}$ \cite{SSantos:2015mzs}.
Following Schwinger's approach \cite{Schwinger:1951nm}, we use the identity 
\begin{align}
    \label{SPT}
    \ln O=-\int_{0}^{\infty}\frac{ds}{s}e^{-iOs}\,,
\end{align}
in \eqref{L2} to obtain
\begin{align}
    \label{L3}
    \mathcal{L}^{\left(1\right)}&=\frac{i}{2}tr\int_{0}^{\infty}\frac{ds}{s}\left(e^{-i\mathcal{H}s}-e^{-i\mathcal{H}_{0}s}\right)\,.
\end{align}
According to \cite{Schwinger:1951nm}, the parameter $s$ is identified as the proper-time. So that we can interpret the exponentials in \eqref{L3} as time evolution operators which are responsible to perform the following transformations
\begin{eqnarray}
   \label{evo}
    B\left(s\right)&=& e^{i\mathcal{H}s}B\left(0\right)e^{-i\mathcal{H}s}\;,\nonumber\\
    \left|x'\left(s\right)\right>&=&e^{-iHs}\left|x'\left(0\right)\right>\,.
\end{eqnarray}
We also define
\begin{eqnarray}
    \label{posi}
    x'\left(s\right)\left|x'\left(s\right)\right>&=&x'\left|x'\left(s\right)\right>\;,\nonumber\\
    x\left(0\right)\left|x\right>&=&x\left|x\right>\;.
\end{eqnarray}
The one-loop effective Lagrangian is then more properly written as
\begin{align}
    \label{L4}
    \mathcal{L}^{\left(1\right)}&=\frac{i}{2}\int_{0}^{\infty}\frac{ds}{s}tr\left[U\left(x',x,s\right)-U_{0}\left(x',x,s\right)\right]\,,
\end{align}
with $U\left(x',x,s\right)=e^{-i\mathcal{H}s}$ and $U_{0}\left(x',x,s\right)=e^{-i\mathcal{H}_{0}s}$. Thus, the problem now resumes to evaluate these time evolution operators. For this purpose, we can interpret $U\left(x',x,s\right)$ as the operator describing the evolution of a quantum mechanical system governed by the ``Hamiltonian'' $\mathcal{H}$ at the ``time'' $s$. In classical mechanics, the temporal evolution operator is responsible for the transformations $x_{\mu}\left(s\right)=x_{\mu}\left(x\left(0\right),p\left(0\right),s\right)$ and $p_{\mu}\left(s\right)=p_{\mu}\left(x\left(0\right),p\left(0\right),s\right)$. Which is found by solving the Hamilton-Jacobi equations \cite{goldstein1950classical}. In this case, we are dealing with a quantum system, the equations corresponding to the Hamilton-Jacobi equations for such a system are given by \cite{dirac1958principles,Fock:1937dy,Nambu:1950rs}
\begin{widetext}
\begin{eqnarray}
    \label{QHJ}
    \mathcal{H}U\left(x'\left(s\right),x\left(0\right),s\right)&=&i\frac{\partial}{\partial s}U\left(x'\left(s\right),x\left(0\right),s\right)\;,\nonumber\\
    \Pi^{\mu}\left(0\right)U\left(x'\left(s\right),x\left(0\right),s\right)&=&\left(-i\partial^{\mu}-eA^{\mu}\left(x\right)\right)U\left(x'\left(s\right),x\left(0\right),s\right)\;,\nonumber\\
    \Pi^{\mu}\left(s\right)U\left(x'\left(s\right),x\left(0\right),s\right)&=&\left(i\partial'^{\mu}-eA^{\mu}\left(x'\right)\right)U\left(x'\left(s\right),x\left(0\right),s\right)\,,
\end{eqnarray}
\end{widetext}
with the initial condition
\begin{align}
    \label{Condiinicial}
    \lim_{s\rightarrow0}U\left(x',x,s\right)&=\delta\left(x'-x\right)\,.
\end{align}
Thus, to find $U\left(x',x,s\right)$, one needs to solve the set of differential equations in \eqref{QHJ} with the initial condition \eqref{Condiinicial}, together with the Heisenberg equations of motion,
\begin{eqnarray}
    \label{EqMov}
    \frac{dx_{\mu}\left(s\right)}{ds}&=&-i\left[x_{\mu}\left(s\right),\mathcal{H}\right]\;,\nonumber\\
    \frac{d\Pi_{\mu}\left(s\right)}{ds}&=&-i\left[\Pi_{\mu}\left(s\right),\mathcal{H}\right]\,.
\end{eqnarray}
This task can only be achieved if the fields are known.

Using the procedure described above, we will now apply to the case where the fields $\mathcal{F}^{\mu\sigma}$, $b$ and $m_5$ are constant in space and time. So, the first Hamiltonian in \eqref{H} becomes
\begin{align}
    \label{H2}
    \mathcal{H}&=\Pi^{2}+N^{\nu}\Pi_{\nu}+\mathcal{M}
\end{align}
with
\begin{eqnarray}
    \label{HDes}
    \mathcal{M}&=&-\frac{e}{2}\sigma_{\mu\nu}\mathcal{F}^{\mu\nu}+\left(b^{2}-m_{5}^{2}\right)-2ibm_{5}\gamma_{0}-2imm_{5}\gamma_{5}+m^{2}\;,\nonumber\\
    N^{\nu}&=&\left(2im_{5}\gamma_{5}\gamma^{\nu}-2ib\gamma_{5}\sigma_{0}^{\,\,\nu}\right)\,.
\end{eqnarray}
In the case of constant fields, the Heisenberg equations of motion \eqref{EqMov} become
\begin{eqnarray}
    \label{EqMov2}
    \frac{d\Pi_{\mu}\left(s\right)}{ds}&=&e\mathcal{F}_{\mu\nu}\left(2\Pi^{\nu}\left(s\right)+N^{\nu}\right)\;,\nonumber\\
    \frac{dx_{\mu}\left(s\right)}{ds}&=&2\Pi_{\mu}\left(s\right)+N_{\mu}\,.
\end{eqnarray}
These equations can be solved for  $\Pi_{\mu}\left(s\right)$ and $\Pi_{\mu}\left(0\right)$ in terms of $x_{\mu}\left(s\right)$ and $x_{\mu}\left(0\right)$,
\begin{eqnarray}
    \label{Moment}
    \Pi_{\mu}\left(s\right)&=&\frac{1}{2}e^{e\mathcal{F}_{\mu}^{\,\,\alpha}s}\left(\frac{e\mathcal{F}}{\sinh\left(e\mathcal{F}s\right)}\right)_{\alpha}^{\,\,\beta}\left[x_{\beta}\left(s\right)-x_{\beta}\left(0\right)\right]+\nonumber\\
    &-&\frac{1}{2}N_{\mu}\;,\nonumber\\
    \Pi_{\mu}\left(0\right)&=&\left(\frac{e\mathcal{F}}{e^{2e\mathcal{F}s}-1}\right)_{\mu}^{\,\,\beta}\left[x_{\beta}\left(s\right)-x_{\beta}\left(0\right)\right]-\frac{1}{2}N_{\mu}\,.
\end{eqnarray}
Going back to \eqref{H2}, we get
\begin{eqnarray}
    \label{H4}
    \mathcal{H}&=&\left[x_{\beta}\left(s\right)-x_{\beta}\left(0\right)\right]K^{\beta\gamma}\left[x_{\gamma}\left(s\right)-x_{\gamma}\left(0\right)\right]+\nonumber\\
    &-&\frac{i}{2}tr\left[\frac{e\mathcal{F}}{\tanh\left(e\mathcal{F}s\right)}\right]-\frac{1}{4}N^{\mu}N_{\mu}+\mathcal{M}\,,
\end{eqnarray}
with $K^{\beta\gamma}=\frac{1}{4}\left[\left(\frac{e\mathcal{F}}{\sinh\left(e\mathcal{F}s\right)}\right)^{2}\right]^{\beta\gamma}$. Now, the first equation in \eqref{QHJ} can be solved by using \eqref{H4}. Thus,
\begin{widetext}
\begin{align}
    \label{U1}
    U\left(x',x,s\right)&=\frac{C\left(x',x\right)}{s^{2}}e^{\frac{i}{4s}\left[x^{\mu}\left(s\right)-x^{\mu}\left(0\right)\right]\left[\frac{e\mathcal{F}s}{\tanh\left(e\mathcal{F}s\right)}\right]_{\mu\nu}\left[x^{\nu}\left(s\right)-x^{\nu}\left(0\right)\right]}e^{-\frac{1}{2}tr\ln\frac{\sinh\left(e\mathcal{F}s\right)}{e\mathcal{F}s}}e^{-i\mathcal{Q}s}\,,
\end{align}
\end{widetext}
where $\mathcal{Q}=-\frac{1}{4}N^{\mu}N_{\mu}+\mathcal{M}$. The object $C\left(x',x\right)$ is an integration constant
and has no explicit dependency on $s$. Using the remaining equations in \eqref{QHJ} and \eqref{Moment}, we can find an expression for $C\left(x',x\right)$, which is
\begin{align}
    \label{C2}
    C\left(x',x\right)&=C_{0}e^{-i\int_{x}^{x'}\left(eA^{\mu}-\frac{1}{2}N^{\mu}-\frac{1}{2}e\mathcal{F}^{\mu\beta}\left[y_{\beta}\left(s\right)-y_{\beta}\left(0\right)\right]\right)dy}\,,
\end{align}
where $C_{0}$ is determined using the initial condition \eqref{Condiinicial},
\begin{align}
    \label{C0}
    C_{0}&=-\frac{i}{\left(4\pi\right)^{2}}\,.
\end{align}
Thence,
\begin{widetext}
\begin{align}
    \label{U2}
    U\left(x',x,s\right)&=-\frac{i}{\left(4\pi s\right)^{2}}e^{\frac{i}{4s}\left[x^{\mu}\left(s\right)-x^{\mu}\left(0\right)\right]\left[\frac{e\mathcal{F}s}{\tanh\left(e\mathcal{F}s\right)}\right]_{\mu\nu}\left[x^{\nu}\left(s\right)-x^{\nu}\left(0\right)\right]}e^{-\frac{1}{2}tr\ln\frac{\sinh\left(e\mathcal{F}s\right)}{e\mathcal{F}s}}e^{-i\mathcal{Q}s}\,.
\end{align}
\end{widetext}
To find an expression for the $U_{0}\left(x,x,s\right)$, we can just set $\mathcal{F}_{\mu\nu}$, $b$, and $m_5$ to zero in \eqref{U2}, resulting in
\begin{align}
    \label{U02}
    U_{0}\left(x',x,s\right)&=-\frac{i}{\left(4\pi s\right)^{2}}e^{-im^{2}s}\,.
\end{align}
Now, since we are interested in the vacuum-vacuum transition amplitude, we have to take the limit $x'=x$. Thus, \eqref{U2} and \eqref{U02} become
\begin{eqnarray}
    \label{Us}
    U\left(x,x,s\right)&=&-\frac{i}{\left(4\pi s\right)^{2}}e^{-im^{2}s}e^{-\frac{1}{2}tr\ln\frac{\sinh\left(e\mathcal{F}s\right)}{e\mathcal{F}s}}e^{-i\mathcal{Q}'s}\;,\nonumber\\
    U_{0}\left(x,x,s\right)&=&-\frac{i}{\left(4\pi s\right)^{2}}e^{-im^{2}s}\,,
\end{eqnarray}
with
\begin{align}
    \label{Q'}
    \mathcal{Q}'&=-\frac{e}{2}\sigma_{\mu\nu}\mathcal{F}^{\mu\nu}+\left(4b^{2}-3m_{5}^{2}\right)+4ibm_{5}\gamma_{0}-2imm_{5}\gamma_{5}\,.
\end{align}
The effective Lagrangian is then rewritten as
\begin{eqnarray}
    \label{L6}
    \mathcal{L}^{\left(1\right)}&=&\frac{1}{32\pi^{2}}\int_{0}^{\infty}\frac{ds}{s^{3}}e^{-im^{2}s}\cdot\nonumber\\
    &\cdot&\left[e^{-\frac{1}{2}tr\ln\frac{\sinh\left(e\mathcal{F}s\right)}{e\mathcal{F}s}}tr\left(e^{-i\mathcal{Q}'s}\right)-4\right]\,.
\end{eqnarray}
The problem to evaluate the effective Lagrangian is reduced to the problem of finding the eigenvalues of $\mathcal{F}_{\mu\nu}$ and $\mathcal{Q}'$. The first one was already calculated by \cite{Schwinger:1951nm} and is given by
\begin{align}
    \label{trace1}
    \exp\left[-\frac{1}{2}tr\ln\frac{\sinh\left(e\mathcal{F}s\right)}{e\mathcal{F}s}\right]&=\left[\frac{\left(es\right)^{2}ab}{\sinh\left(ebs\right)\sin\left(eas\right)}\right]\,,
\end{align}
where
\begin{align}
    \label{ab}
    a&=\left(\sqrt{\mathcal{Y}^{2}+\mathcal{R}^{2}}+\mathcal{Y}\right)^{\frac{1}{2}},\qquad b=\left(\sqrt{\mathcal{Y}^{2}+\mathcal{R}^{2}}-\mathcal{Y}\right)^{\frac{1}{2}},
\end{align}
with $\mathcal{R}=\frac{1}{4}\mathcal{F}^{\mu\sigma}\widetilde{\mathcal{F}}_{\mu\sigma}=\mathcal{E}\cdot\mathcal{B}$ being the Pontryagin form and $\mathcal{Y}=\frac{1}{4}\mathcal{F}_{\lambda\sigma}\mathcal{F}^{\lambda\sigma}=\frac{1}{2}\left(\mathcal{\mathcal{B}}^{2}-\mathcal{E}^{2}\right)$ being the Maxwell form.

Because our final goal is to compute the E-P pair creation probability, we proceed with one more simplification, the pure electric field approximation. Moreover, according to \cite{Schwinger:1951nm,Dittrich:1978fc,Valluri:1982ip}, the probability of pair creation within constant background fields only occurs for pure electric field. Therefore, writing $\mathcal {Q}'$ in its explicit matrix form and taking the pure electric field limit we find the following eigenvalues,
\begin{align}
    \label{AutovaloresQ}
    \left(\mathcal{Q}'\right)^{EV}&=\alpha\pm\sqrt{\left(4ibm_{5}\right)^{2}-\left(e\mathcal{E}\pm2mm_{5}\right)^{2}}\,.
\end{align}
where $\alpha=\left(4b^{2}-3m_{5}^{2}\right)$.
Using \eqref{trace1} and \eqref{AutovaloresQ}, we obtain the following expression for the pure electric field effective Lagrangian \begin{widetext}
\begin{align}
    \label{L7}
    \mathcal{L}_{\left(\mathcal{E}\right)}^{\left(1\right)}&=\frac{1}{8\pi^{2}}\int_{0}^{\infty}\frac{ds}{s^{3}}e^{-im^{2}s}\left\{ e^{is\alpha}e\mathcal{E}s\frac{\cosh\left[\frac{1}{2}s\left(a_{+}^{\left(\mathcal{E}\right)}+a_{-}^{\left(\mathcal{E}\right)}\right)\right]\cosh\left[\frac{1}{2}s\left(a_{+}^{\left(\mathcal{E}\right)}-a_{-}^{\left(\mathcal{E}\right)}\right)\right]}{\sinh\left(e\mathcal{E}s\right)}-1\right\}\,,
\end{align}
\end{widetext}
with
\begin{eqnarray}
    \label{aa}
    a_{+}^{\left(\mathcal{E}\right)}&=&\sqrt{\beta^{2}+\left(e\mathcal{E}-i\delta\right)^{2}}\;,\nonumber\\
    a_{-}^{\left(\mathcal{E}\right)}&=&\sqrt{\beta^{2}+\left(e\mathcal{E}+i\delta\right)^{2}}\;,\nonumber\\
    \beta&=&4bm_{5}\;,\nonumber\\
    \delta&=&2mm_{5}\,.
\end{eqnarray}

In order to complement the analysis of the effective Lagrangian \eqref{L7}, it is necessary to establish whether it has divergent terms when $s\to0$. To verify it, we expand the first term inside of braces in \eqref{L7} in a power series of $s$. The divergent terms are then subtracted due renormalization: The electric field divergent part is combined with the free part of \eqref{L0} \cite{Schwinger:1951nm}; The Lorentz-violating part is absorbed in a similar way as in \cite{Sitenko:2002fy}. Finally, we have the fully renormalized effective Lagrangian for the pure electric field
\begin{widetext}
\begin{eqnarray}
    \label{L8}
    \mathcal{L}_{\left(\mathcal{E}\right)}&=&\frac{1}{2}\mathcal{E}^{2}+\frac{1}{8\pi^{2}}\int_{0}^{\infty}\frac{ds}{s^{3}}e^{-im^{2}s}\left(e^{is\alpha} e\mathcal{E}s\frac{\cosh\left[\frac{1}{2}s\left(a_{+}^{\left(\mathcal{E}\right)}+a_{-}^{\left(\mathcal{E}\right)}\right)\right]\cosh\left[\frac{1}{2}s\left(a_{+}^{\left(\mathcal{E}\right)}-a_{-}^{\left(\mathcal{E}\right)}\right)\right]}{\sinh\left(e\mathcal{E}s\right)}+\right.\nonumber\\
    &-&\left.is\alpha+\frac{1}{12}s^{2}\left[6\left(\alpha^{2}+\beta^{2}+\delta^{2}\right)-4\left(e\mathcal{E}\right)^{2}\right]-1\right)\,.
\end{eqnarray}
\end{widetext}
One can easily check that at the limit of vanishing Lorentz-violating fields in \eqref{L8}, the usual result of \cite{Schwinger:1951nm} is recovered.

It is also interesting and useful to consider the expansion of \eqref{L8} to the first order in the interaction between the fields. In this case, one gets
\begin{align}
    \label{L10}
    \mathcal{L}_{\left(\mathcal{E}\right)}^{\left(1\right)}&=\frac{e^{2}\mathcal{E}^{2}}{6m^{2}\pi^{2}}\left(b^{2}-\frac{3}{4}m_{5}^{2}\right)\,.
\end{align}
One readily notices that an CPT-even term is generated, a result previously encountered in \cite{Ferrari:2021eam,Gomes:2009ch,BaetaScarpelli:2013rmt}.

\section{Schwinger Effect}\label{PC}

In this section we finally compute at the probability of creating an E-P pair due to the external field (Lorentz-violating and Electric), which is known as the Schwinger effect. The probability of creating an E-P pair is given by the following relation \cite{dittrich1985effective2}
\begin{align}
    \label{prob}
    P&=1-\left|\left\langle 0_{+}|0_{-}\right\rangle \right|^{2}=1-e^{-2ImW\left[A\right]}\,,
\end{align}
where $W\left[A\right]=\int d^{4}x\;\mathcal{L}$. Taking small values of $W\left[A\right]$, the probability can be approximated for \cite{dittrich1985effective2}
\begin{align}
    \label{prob2}
    P&=2\int d^{4}x\;Im\mathcal{L}\,.
\end{align}
Thus, the probability density of creating E-P pairs is just $\mathcal{P}=2Im\mathcal{L}$. Therefore, to obtain the E-P pair creation probability, we need only the imaginary part of the effective Lagrangian\footnote{The essence of the computation from Equation \eqref{L8} to Equation \eqref{IL4} follow \cite{dittrich1985effective2}.}. Taking the advantage that the imaginary part of
\begin{align}
    e^{ix}=\cos{x}+i\sin{x}\;,
\end{align}
is an odd function and the real part is even, the imaginary part of the effective Lagrangian reads
\begin{widetext}
\begin{align}
\label{ILL}
    Im\mathcal{L}_{\left(\mathcal{E}\right)}^{\left(1\right)}&=\frac{1}{8\pi^{2}}\int_{0}^{\infty}\frac{ds}{s^{3}}\left[\left(\sin\left[s\left(\alpha-m^{2}\right)\right]e\mathcal{E}s\frac{\cosh\left[\frac{1}{2}s\left(a_{+}^{\left(\mathcal{E}\right)}+a_{-}^{\left(\mathcal{E}\right)}\right)\right]\cosh\left[\frac{1}{2}s\left(a_{+}^{\left(\mathcal{E}\right)}-a_{-}^{\left(\mathcal{E}\right)}\right)\right]}{\sinh\left(e\mathcal{E}s\right)}\right)\right.\nonumber\\&\left.-s\alpha\cos\left(m^{2}s\right)-\sin\left(m^{2}s\right)\left(\frac{1}{12}s^{2}\left[6\left(\alpha^{2}+\beta^{2}+\delta^{2}\right)-4\left(e\mathcal{E}\right)^{2}\right]-1\right)\right]\;,
\end{align}
\end{widetext}
and then, rearranging the terms using the fact that
\begin{align}
    \sin{x}=\frac{1}{2i}\left(e^{ix}-e^{-ix}\right)\;,\nonumber\\
    \cos{x}=\frac{1}{2}\left(e^{ix}+e^{-ix}\right)\;,
\end{align}
the imaginary part of \eqref{L8} yields
\begin{widetext}
\begin{align}
    \label{IL}
    Im\mathcal{L}_{\left(\mathcal{E}\right)}^{\left(1\right)}&=\frac{1}{8\pi^{2}}\frac{1}{2i}\int_{-\infty}^{\infty}\frac{ds}{s^{3}}e^{-im^{2}s}\left\{e^{is\alpha}e\mathcal{E}s\frac{\cosh\left[\frac{1}{2}s\left(a_{+}^{\left(\mathcal{E}\right)}+a_{-}^{\left(\mathcal{E}\right)}\right)\right]\cosh\left[\frac{1}{2}s\left(a_{+}^{\left(\mathcal{E}\right)}-a_{-}^{\left(\mathcal{E}\right)}\right)\right]}{\sinh\left(e\mathcal{E}s\right)}\right.\nonumber\\&-\left.is\alpha+\frac{1}{12}s^{2}\left[6\left(\alpha^{2}+\beta^{2}+\delta^{2}\right)-4\left(e\mathcal{E}\right)^{2}\right]-1\right\}\,.
\end{align}
\end{widetext}
In comparison with \eqref{ILL} we have changed the limits of the integral, and now we can proceed and solve equation \eqref{IL} employing the residue method. The poles appear as $s_{n}=-\frac{in\pi}{e\mathcal{E}}$ due to the function $\sinh$ in the denominator. Closing the contour by a semi-circle in the lower half-plane, in order to guarantee that we will have a pair creation rate with a decreasing exponential, and computing the residues of the simple poles using the Laurent series of the function $\frac{1}{\sinh\left(e\mathcal{E}s\right)}$, we have
\begin{widetext}
\begin{eqnarray}
    \label{IL2}
    Im\mathcal{L}_{R}^{\left(1\right)}\left(\mathcal{E}\right)&=&\frac{\left(e\mathcal{E}\right)^{2}}{8\pi^{3}}\sum_{n=1}^{\infty}\frac{\left(-1\right)^{n}}{n^{2}}\cosh\left[\frac{1}{2}\frac{in\pi}{e\mathcal{E}}\left(\sqrt{\left(4bm_{5}\right)^{2}+\left(e\mathcal{E}+2mm_{5}\right)^{2}}+\sqrt{\left(4bm_{5}\right)^{2}+\left(e\mathcal{E}-2mm_{5}\right)^{2}}\right)\right]\cdot\nonumber\\
    &\cdot&\cosh\left[\frac{1}{2}\frac{in\pi}{e\mathcal{E}}\left(\sqrt{\left(4bm_{5}\right)^{2}+\left(e\mathcal{E}+2mm_{5}\right)^{2}}-\sqrt{\left(4bm_{5}\right)^{2}+\left(e\mathcal{E}-2mm_{5}\right)^{2}}\right)\right]\cdot\nonumber\\
    &\cdot&e^{-\frac{n\pi}{e\mathcal{E}}\left(m^{2}-4b^{2}+3m_{5}^{2}\right)}\,.
\end{eqnarray}
\end{widetext}
Since the fields which are Lorentz-violating are very small, we can expand \eqref{IL2} up to the second order in Lorentz-violating fields and write
\begin{eqnarray}
    \label{IL4}
    Im\mathcal{L}_{R}^{\left(1\right)}\left(\mathcal{E}\right)&=&\sum_{n=1}^{\infty}\frac{\left(e\mathcal{E}\right)^{2}}{8n^{2}\pi^{3}}e^{-\frac{m^{2}n\pi}{e\mathcal{E}}}\left[1-\left(\frac{2mm_{5}n\pi}{e\mathcal{E}}\right)^{2}+\right.\nonumber\\
    &-&\left.\frac{3m_{5}^{2}n\pi}{e\mathcal{E}}+\frac{4n\pi b^{2}}{e\mathcal{E}}\right]\,,
\end{eqnarray}

We end this section by commenting that the contributions of the Lorentz-violating background fields to the rate of E-P pair creation is quite small because the Lorentz-Violating fields are on the order of $10^{-14} \,GeV$ \cite{Kostelecky:2008bfz}. Nevertheless, it could be possible to measure this contribution in recent experiments, especially the LUXE experiment? The next section is devoted to this discussion. Futhermore, we can see that the pseudo-scalar mass $m_5$ indeed contributes for the Schwinger effect, but in square of $m_5$ only, this may be explained by the argument of the chiral rotation only holding at first order in $m_5$, and higher orders could, in principle, appear.

\section{Phenomenology}\label{PHE}

Although the Schwinger effect calculated above is valid for the constant electric field case, experimentally it is still too complicated to create an electric field with this condition \cite{Ringwald:2001ib}. With this problem in hand, some works were done in an attempt to understand if high intensity lasers, which is an alternating electric field, could be used to study the Schwinger effect \cite{Bunkin:1969if,Brezin:1970xf,Popov:1971ff,Popov:1971iga,Popov:1973uw,Marinov:1977gq,Schutzhold:2008pz}. To get around this problem, experiments pretend to use high-intensity lasers to probe this phenomenon, for exemple, SLAC-E144 (which was the first experiment with the goal to explore the Schwinger effect using the nonlinear Compton scattering and Breit-Wheeler process \cite{Altarelli:2019zea}), XFEL, ELI, SEL and LUXE. From an experimental point of view it is easier to obtain high intensities of electromagnetic fields with lasers. Thus, this section is dedicated to discuss how it is possible to relate the constant electric field result to the most realistic situation and if it can be used to probe the Lorentz-violating fields out from the result \eqref{IL4}. Our main focus is the LUXE experiment \cite{Jacobs:2021fbg}.

From the experimental point of view, there are two parameters used to describe the nonlinear processes of QED: The adiabaticity parameter $\gamma=\frac{e\mathcal{E}}{m\omega}$, with $\omega$ being the laser frequency, and the quantum parameter $\xi=\frac{\mathcal{E}}{\mathcal{E}_{cr}}$, which measures the ratio between the background field and the Schwinger critical field. The adiabaticity parameter receives this name because the analogy between vacuum pair production and atomic ionization, which is also used to distinguish the non-perturbative regime from the perturbative one in atomic ionization probability \cite{Dunne:2008kc}. In particular for $\gamma\gg 1$ we have the non-perturbative regime, which is proportional to the result found in Schwinger's work \cite{Schwinger:1951nm,Hartin:2018sha}. Thus, we can approximate our result as a case where $\gamma\gg 1$ and try to compare with the expected result in the LUXE experiment \cite{Jacobs:2021fbg}.

In LUXE experiment, they will use the Breit-Wheeler process to probe the transition from perturbative to the non-perturbative regime of pair creation \cite{Abramowicz:2021zja}. The Breit–Wheeler process is the production of an electron–positron pair from the collision of two photons, for more detailed description of the experimental setup see \cite{Abramowicz:2021zja} and references therein. As said earlier, since the adiabaticity parameter is $\gg 1$ we can approximate the pair creation probability to the result for a constant electric field \cite{Ringwald:2001ib}
\begin{align}
    Im\mathcal{L}_{R}^{\left(1\right)}\left(\mathcal{E}\right)=\frac{\left(e\mathcal{E}\right)^{2}}{8\pi^{3}}\sum_{n=1}^{\infty}\frac{1}{n^{2}}e^{-\frac{n\pi m^{2}}{e\mathcal{E}}}\,.
\end{align}
The central idea is that with the LUXE experiment managing to measure the transition from perturbative to the non-perturbative regime, it can be a useful environment to test the result \eqref{IL4} for, at least, to provide new bounds for the Lorentz-violating parameters. But, in the preliminary studies on the phase $1$, where this transition is intended to be reached, the minimum number of positrons from the E-P creation still does not come close to the value where the Lorentz violation will be relevant. Analyzing \eqref{IL4}, as the values of Lorentz violating fields are very small by current bounds, the predicted precision in the LUXE experiment would not be able to measure it. The meaning of \eqref{IL4} can be given as the probability that an E-P pair is produced per unit time and unit volume \cite{Ringwald:2001ib}. For the Compton space-time volume, $\simeq 10^{-59}m^3s$, that probability to be close to the predicted value for the Lorentz invariant case, the electric field must be of the order $10^{21}\mathcal{E}_{cr}$.\footnote{Although we used the natural units during the calculations, to obtain this prediction for the electric field, the SI units were used and the necessary restoration of the units in the result \eqref{IL4} was done.}, an extremely high value even for the most recent experiments.

Another analysis we can do is to look at the last two terms in \eqref{IL4} and note that depending on the numerical values of the fields $b$ and $m_5$, we may have a small increase ($b>m_5$) in the probability of creating pairs or a decrease ($b<m_5$). Despite being extremely small, we can notice an influence of Lorentz-violating background fields. We can also notice that, if $b=\frac{3}{4}m_5$, the contribution of Lorentz-violating fields vanishes. Therefore, if future experiments cannot measure these contributions, we can have a constraint between the values fields $b$ and $m_5$.

\section{Conclusions}\label{FINAL}

In this paper, we studied the Schwinger effect in the presence of Lorentz-violating background fields. First we calculated the one-loop effective Lagrangian for the pure constant electric field case in the presence of the time-like $b_\mu$ and the pseudo-scalar $m_5$ Lorentz-violating background fields using the proper-time method. An interest result deduced from the effective Lagrangian \eqref{L8} is the emergence of the CPT-even term \eqref{L10}, also found in the following works \cite{Ferrari:2021eam,Gomes:2009ch,BaetaScarpelli:2013rmt} and the contribution of the pseudo-scalar $m_5$ for the Schwinger effect, even though it can be removed by a chiral transformation from the classical action \cite{Kostelecky:2013rta}, as mentioned before, this could be explained by the fact that the chiral transformation only holds in first order in $m_5$ at quantum level.

The contribution of Lorentz-violating background fields contribute for the Schwinger effect is displayed in expression \eqref{IL4}, which is the leading contribution to the Schwinger effect for a constant electric field. We found, as expected, that the contributions are very small. Therefore, a high precision in the measurements of E-P pairs is necessary if we want to test the violation of Lorentz symmetry out from the Schwinger effect. Furthermore, it is necessary that the experiment is able to investigate the non-perturbative regime of pair creation so that it is also possible to test the result \eqref{IL4}. This goal was proposed by the LUXE experiment, which should be installed by the year 2024. But even so, the predicted precision for phase-1 would still not be enough to measure the contribution of Lorentz-violating fields. In fact, we found that we may have a small increase in the creation probability if $b>m_5$ or a decrease if $b<m_5$. A possible constraint between the fields $b$ and $m_5$ can also be inferred if future experiments cannot measure these contributions.

We conclude by commenting that other Lorentz-violating fields can also be taken into account in the Schwinger effect. The analysis of how they would contribute to the phenomenon still lacks. Likewise, one can also take into account the dynamic Schwinger effect in the presence of these Lorentz-violating background fields.

\section*{Acknowledgements}

This study was financed in part by The Coordena\c c\~ao de Aperfei\c coamento de Pessoal de N\'ivel Superior - Brasil (CAPES) - Finance Code 001.

\bibliography{library}
\bibliographystyle{utphys2}

\end{document}